\documentclass[prl,superscriptaddress,twocolumn]{revtex4}
\pdfoutput=1
\usepackage{amsmath}
\usepackage{xcolor}
\usepackage{graphicx}
\usepackage{bm}
\usepackage{url}
\usepackage{hyperref}

\DeclareMathOperator{\tr}{tr}

\begin{document}

\title{Defect Annihilation and Proliferation in Active Nematics}

\author{Luca Giomi}
\affiliation{SISSA, International School for Advanced Studies, Via Bonomea 265, 34136 Trieste, Italy}
\author{Mark J. Bowick}
\affiliation{Physics Department, Syracuse University, Syracuse NY 13244, USA}
\affiliation{Syracuse Biomaterials Institute, Syracuse University, Syracuse NY 13244, USA}
\author{Xu Ma}
\affiliation{Physics Department, Syracuse University, Syracuse NY 13244, USA}
\author{M.~Cristina Marchetti}
\affiliation{Physics Department, Syracuse University, Syracuse NY 13244, USA}
\affiliation{Syracuse Biomaterials Institute, Syracuse University, Syracuse NY 13244, USA}

\begin{abstract}
Liquid crystals inevitably possess topological defect excitations generated through boundary conditions, through
applied fields, or in quenches to the ordered phase. In equilibrium, pairs of defects coarsen and annihilate as the uniform ground state 
is approached. Here we show that defects in active liquid crystals exhibit profoundly different behavior, depending on the degree of 
activity and its contractile or extensile character. While contractile systems enhance the annihilation dynamics of passive systems, 
extensile systems act to drive defects apart so that they swarm around in the manner of topologically well-characterized self-propelled 
particles. We develop a simple analytical model for the defect dynamics which reproduces the key features of both the numerical solutions 
and recent experiments on microtubule-kinesin assemblies. 
\end{abstract}

\maketitle
Active liquid crystals are nonequilibrium fluids composed of internally driven elongated units. 
The key feature that distinguishes them from their well-studied passive counterparts is that they are maintained 
out of equilibrium not by an external force applied at the system's boundary, such as an imposed shear, but by an 
energy input on each individual unit. Examples include mixtures of cytoskeletal filaments and associated motor 
proteins, bacterial suspensions, the cell cytoskeleton, and even nonliving analogues, such as monolayers of 
vibrated granular rods \cite{Marchetti:2013}. The internal drive that characterizes active liquid crystals dramatically 
changes the system's dynamics and yields novel effects arising from the interplay of orientational order and flow, 
such as spontaneous laminar flow \cite{Voituriez:2005,Marenduzzo:2007,Giomi:2008}, large density 
fluctuations \cite{Ramaswamy:2003,Mishra:2006,Narayan:2007}, unusual rheological 
properties \cite{Sokolov:2009,Giomi:2010,Fielding:2011}, excitability \cite{Giomi:2011,Giomi:2012} and low Reynolds 
number turbulence \cite{Giomi:2012,Wensink:2012}.

Ordered liquid crystalline phases of active matter, like their equilibrium counterparts, exhibit distinctive inhomogeneous 
configurations known as topological defects. In equilibrium, defect configurations may be generated through boundary conditions, 
externally applied fields, or rapid quenches to the ordered state. When the constraints are removed or the system is given time 
to equilibrate, the defects ultimately annihilate~\cite{Kleman:2003}. Experiments have shown that in active systems, in contrast, 
defect configurations can occur spontaneously in bulk and be continuously regenerated by the local energy 
input~\cite{Kemkemer:2000,Sanchez:2012}. The nature of the topological defects is, of course, intimately related to the symmetry of 
the system, which can be either polar (like in ferromagnets) or nematic. While the nature of the charge $+1$ defects that occur in 
polar active systems has been studied for some time \cite{Kruse:2004,Kruse:2006,Voituriez:2006,Elgetti:2011}, the properties of 
defects in apolar or nematic active media are still largely unexplored. In these systems the defects are charge $\pm1/2$ 
disclinations \cite{DeGennes:1993}. Such defects have been identified in monolayers of vibrated granular rods~\cite{Narayan:2007} 
and also in active nematic gels assembled {\em in vitro} from microtubules and kinesins. In the latter case the defects were shown 
to drive spontaneous flows in bulk~\cite{Sanchez:2012,Marchetti:2012}. When confined at an oil-water interface, furthermore, the gel 
forms a two-dimensional active nematic film, with self-sustained flows resembling cytoplasmic streaming and the continuous creation and 
annihilation of defect pairs \cite{Sanchez:2012}. 

In this Letter, we examine the effect of activity on the dynamics of disclinations in a two-dimensional nematic 
liquid crystalline film~\cite{foot}. Hydrodynamics plays an important role in controlling the dynamics of defects in liquid crystals.
As the defect moves, the coupling between the orientational order parameter and the flow velocity of the fluid yields what is usually 
called the {\em backflow} which significantly modifies defect dynamics~\cite{Denniston:1996,Toth:2002,Kats:2002,Svensek:2002,Sonnet:2009}. 
Here we show that active stresses dramatically affect the defect dynamics by altering the backflow in such a way as to speed up, slow down,
or even suppress pair annihilation, according to the extent of activity and the typical time scale of orientational relaxation of the 
nematic phase. Moreover, when the latter is very large compared to the time scale associated with activity, relaxation is overwhelmed 
 entirely, leading to defect proliferation. 

\begin{figure}
\includegraphics[width=1\columnwidth]{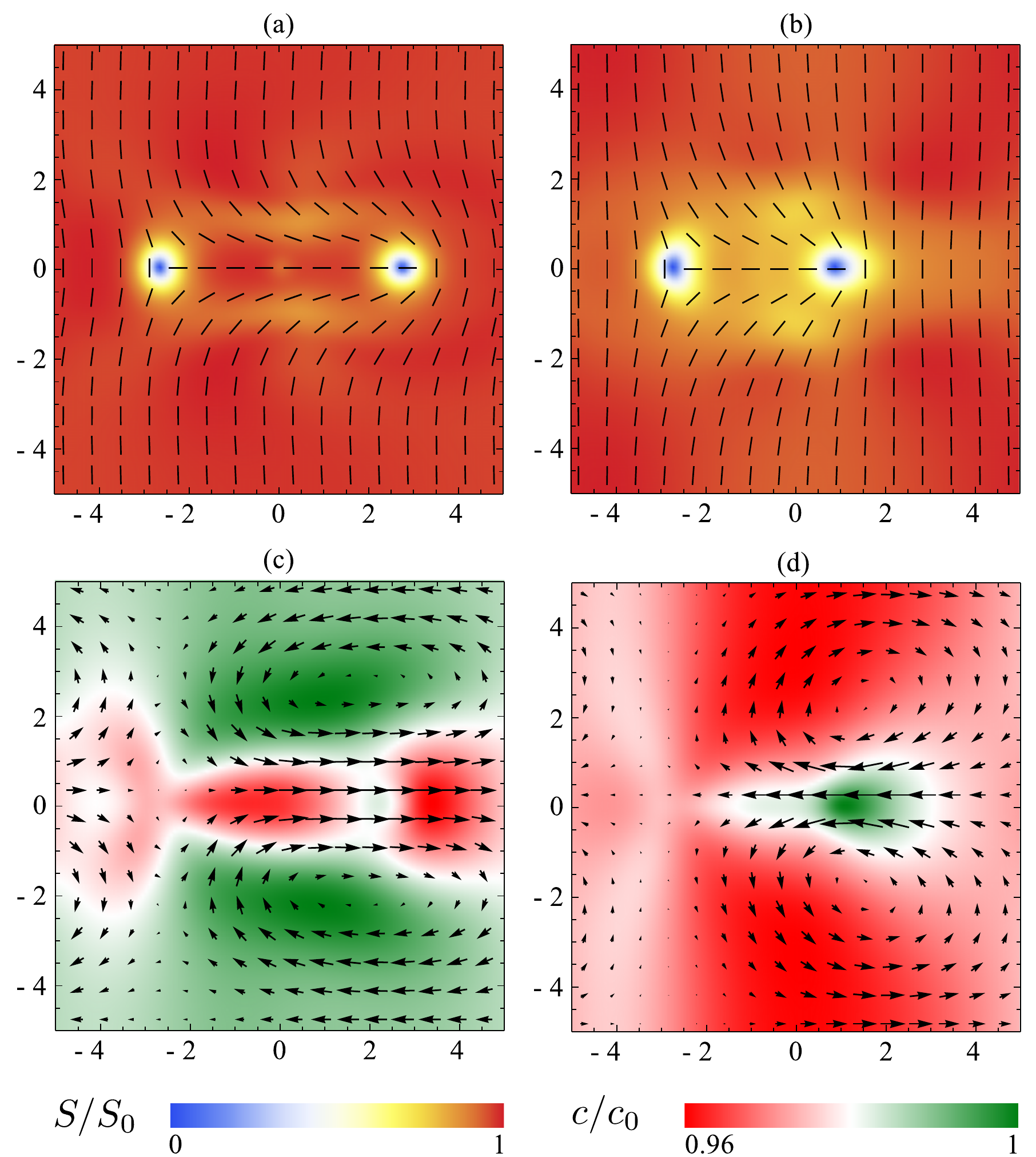}
\caption{(color online) Snapshots of a disclination pair shortly after the beginning of relaxation. (Top) Director field (black lines) 
superimposed on a heat map of the nematic order parameter and (bottom) flow field (arrows) superimposed on a heat map of the concentration 
for an extensile system with $\alpha=-0.2$ (a),(c) and a contractile system with $\alpha=0.2$ (b),(d). In the top images. the color denotes 
the magnitude of the nematic order parameter $S$ relative to its equilibrium value $S_{0}=\sqrt{1-c^{\star}/c_{0}}$. In the bottom images, 
the color denotes the magnitude of the concentration $c$ relative to the average value $c_0$. Depending on the sign of $\alpha$, the backflow 
tends to speed up $(\alpha>0)$ or slow down ($\alpha<0$) the annihilation process by increasing or decreasing the velocity of the $+1/2$ 
disclination. For $\alpha$ negative and sufficiently large in magnitude, the $+1/2$ defect reverses its direction of motion (c) and escapes  
annihilation.}
\label{fig:flows}
\end{figure}

The hydrodynamic equations of active nematic liquid crystals can be obtained from that of passive nematics by the addition of 
nonequilibrium stresses and currents due to activity \cite{Giomi:2011,Giomi:2012,Marchetti:2013}. These equations are formulated in terms of 
a concentration $c$, a flow velocity $\bm{v}$ and the nematic tensor order parameter $Q_{ij}=S\left(n_in_j-\frac12\delta_{ij}\right)$, 
with $\bm{n}$ the director field. The alignment tensor $Q_{ij}$ is traceless and symmetric, and, hence, has only two independent 
components in two dimensions. Considering for simplicity the case of an incompressible fluid of constant density $\rho$, where 
$\nabla\cdot\bm{v}=0$, the equations are given by
\begin{subequations}\label{eq:hydrodynamics}
\begin{gather}
\frac{Dc}{Dt}=\partial_i\left[D_{ij}\partial_jc+\alpha_{1}c^{2}\partial_jQ_{ij}\right]\;,\\[7pt]
\rho \frac{Dv_i}{Dt}=\eta{\color{black} \nabla^2} v_i-\partial_ip+\partial_j\sigma_{ij}\;,\\[5pt]
\frac{D{Q}_{ij}}{Dt}=\lambda Su_{ij}+Q_{ik}\omega_{kj}-\omega_{ik}Q_{kj}+\gamma^{-1}H_{ij}\;,
\end{gather}
\end{subequations}
where $\frac{D}{Dt}=\partial_{t}+\bm{v}\cdot\nabla$ indicates the material derivative, $D_{ij}=D_0\delta_{ij}+D_1Q_{ij}$ is the anisotropic 
diffusion tensor, $\eta$ is the viscosity, $p$ is the pressure, and $\lambda$ is the nematic alignment parameter. 
Here $u_{ij}=(\partial_iv_j+\partial_jv_i)/2$ and $\omega_{ij}=(\partial_iv_j-\partial_jv_i)/2$ are the symmetrized rate of strain tensor 
and the vorticity, respectively. The molecular field $H_{ij}$ embodies the relaxational dynamics of the nematic phase (with $\gamma$ a rotational viscosity) 
and can be obtained from the variation of the Landau-De Gennes free energy of a two-dimensional nematic \cite{DeGennes:1993}, 
$H_{ij}=-\delta F/\delta Q_{ij}$, with
\begin{equation}
F/K=\int dA\,\left[\tfrac{1}{4}(c-c^{\star})\tr\bm{Q}^{2}+\tfrac{1}{4}c(\tr\bm{Q}^2)^{2}+\tfrac{1}{2}|\nabla\bm{Q}|^{2}\right]\;,
\end{equation}
where $K$ is an elastic constant with dimensions of energy, $\tr\bm{Q}^{2}=S^{2}/2$ and $c^{\star}$ is the critical concentration for the 
isotropic-nematic transition, so that, at equilibrium, $S=\sqrt{1-c^{\star}/c}$. Finally, the stress tensor 
$\sigma_{ij}=\sigma^{\rm r}_{ij}+\sigma^{\rm a}_{ij}$ is the sum of the elastic stress due to nematic elasticity, 
$\sigma^{\rm r}_{ij}=-\lambda S H_{ij}+Q_{ik}H_{kj}-H_{ik}Q_{kj}$, where for simplicity we have neglected the Eriksen stress, 
and an active contribution, $\sigma^{\rm a}_{ij}=\alpha_{2}c^{2}Q_{ij}$, which describes contractile or extensile stresses exerted by the 
active particles in the direction of the director field.  In addition, activity yields a curvature-induced current 
$\bm{j}^{\rm a}=-\alpha_{1}c^{2}\nabla\cdot\bm{Q}$ in Eq. (\ref{eq:hydrodynamics}a) that drives units from regions populated by fast-moving 
particles to regions of slow-moving particles. The $c^{2}$ dependence of the active stress and current is appropriate for systems where 
activity arises from pair interactions among the filaments via cross-linking motor proteins. The sign of $\alpha_{2}$ depends on whether the 
active particles generate contractile or extensile stresses, with $\alpha_{2}>0$ for the contractile case and $\alpha_{2}<0$ for extensile 
systems, while we assume $\alpha_1>0$.  

To study the dynamics of defects, we consider a pair of opposite-sign half-integer disclinations separated by a distance $x=x_{+}-x_{-}$, 
where $x_{\pm}$ is the $x$ coordinate of the $\pm 1/2$ disclination, respectively, as shown in Figs. \ref{fig:flows}(a) and 
\ref{fig:flows}(b). When backflow is neglected, the pair dynamics is purely relaxational and is controlled by the balance of the attractive
force between defects $\bm{F}_{\rm pair}=-\nabla E_{\rm pair}$, with $E_{\rm pair}\sim K\log (x/a)$ the energy of a defect pair 
(with $a$ the core radius), and an effective frictional force $\bm{F}_{\rm fric}=\mu\dot{\bm{x}}$, with $\mu\sim\gamma$ a friction 
coefficient. Thus $\mu \dot{x}= K/x$ and the distance between the annihilating defects decreases according to a square-root law, 
$x(t)\propto\sqrt{t_{\rm a}-t}$, with $t_{\rm a}$ the annihilation time. More precise calculations have shown that the effective friction 
is itself a function of the defect separation \cite{Pleiner:1988,Ryskin:1991}, $\mu=\mu_{0}\log (x/a)$, although this does not imply 
substantial changes in the overall picture. This simple model predicts that the defect and antidefect approach each other along 
symmetric trajectories. 

We have integrated numerically Eqs. \eqref{eq:hydrodynamics} for an initial configuration of uniform concentration and zero flow velocity, 
with two disclinations of charge $\pm 1/2$ located on the $x$ axis of a square $L\times L$ box at initial positions 
$\bm{x}_{\pm}(0)=(\pm L/4,0)$. The integration is performed using the finite differences scheme described in 
Ref. \cite{Giomi:2011,Giomi:2012}. To render Eqs. \eqref{eq:hydrodynamics} dimensionless, we normalize distance by the approximate length 
of the active rods $\ell=1/\sqrt{c^{\star}}$, stress by the elastic stress of the nematic phase $\sigma=K\ell^{-2}$ and time by 
$\tau=\eta\ell^{2}/K$ representing the ratio between viscous and elastic stress. In these dimensionless units, for simplicity, we let 
$\alpha_2=\alpha$ and take $\alpha_{1}=|\alpha_{2}|/2$. Periodic boundary conditions are assumed, and the defects are allowed to evolve until
they annihilate. Figure \ref{fig:flows} shows a snapshot of the order parameter and flow field shortly after the beginning of the relaxation
for both a contractile and extensile system, with $\alpha=\pm 0.2$ in the units defined above (see also the supplementary movie 
S1~\cite{Supp}).  

\begin{figure}[t]
\includegraphics[width=\columnwidth]{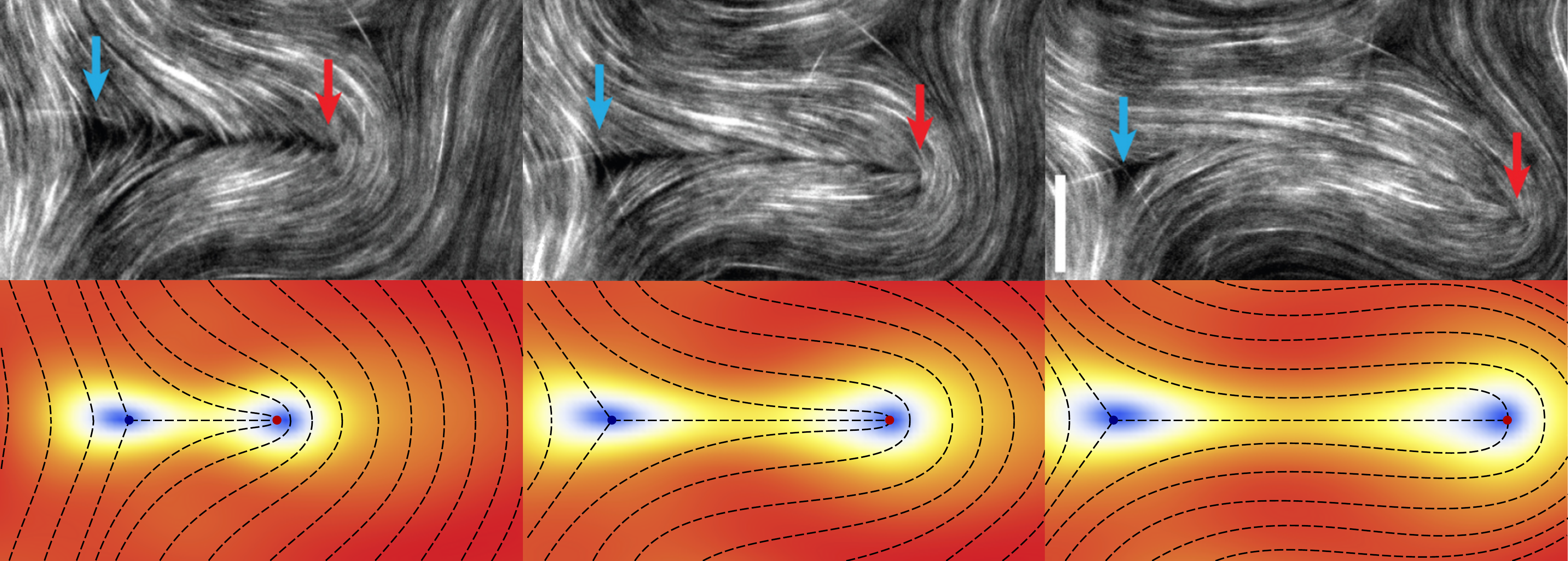}
\caption{\label{fig:pair_production}(color online) Defect pair production in an active suspension of microtubules and kinesin (top) and the 
same phenomenon observed in our numerical simulation of an extensile nematic fluid with $\gamma=100$ and $\alpha=-0.5$. The experimental 
picture is reprinted by permission from T. Sanchez {\emph et al.,} Nature (London) 491, 431 (2012). Copyright 2012, Macmillan.}
\end{figure}

In passive nematic liquid crystals (i.e., $\alpha=0$) it is well known that the dynamics of defects is greatly modified by the so-called 
backflow, that is, the flow induced by reorientation of the nematic order parameter through the elastic stresses $\sigma_{ij}^r$ in the 
Navier-Stokes equation. In particular, when backflow is neglected, the defect and antidefect are predicted to move at the same velocity 
toward each other until annihilation. Backflow tends to speed up the $+1/2$ defect and to slow down the $-1/2$ defect, yielding asymmetric 
trajectories~\cite{Toth:2002}. In active liquid crystals, the active stress in the Navier-Stokes equation provides a new source for flow 
associated with inhomogeneities in the order parameter, as demonstrated first in a one-dimensional thin film geometry where activity drives 
a transition to a spontaneously flowing state \cite{Voituriez:2005}. This new \emph{active backflow} can greatly exceed the curvature-driven 
backflow present in passive systems. Furthermore, the direction of the active backflow is controlled by the sign of the activity parameter 
$\alpha$ and, for a given director configuration, has opposite directions in contractile and extensile systems. Backflow arising from active 
stresses drives the $+1/2$ defect to move in the direction of its ``tail'' in contractile systems ($\alpha>0$) and in the direction of its 
``head'' in extensile systems ($\alpha<0$), where the terminology arises from the cometlike shape of $+1/2$ defects. In contrast, due to 
symmetry considerations, the active backflow vanishes at the core of a $-1/2$ defect which thus remains stationary under the action of active 
stresses. We note that active curvature currents in the concentration equation controlled by $\alpha_1$ have a similar effect, as first noted 
by Narayan, Ramaswamy, and Menon and collaborators in a system of vibrated granular rods~\cite{Narayan:2007}. Such active curvature currents 
control dynamics in systems with no momentum conservation but are very small here, where the concentration variations remain small, as seen 
from Figs. \ref{fig:flows}(c) and \ref{fig:flows}(d), and flow controls the dynamics.

In contractile systems active backflow yields a net {\color{black}speed-up} of the $+1/2$ defect towards its antidefect for the annihilation 
shown in Fig.~\ref{fig:flows}(b). In extensile systems, with $\alpha<0$, backflow drives the $+1/2$ defect to move towards its head, away 
from its $-1/2$ partner in the configuration of Fig. \ref{fig:flows}(b), acting like an effectively {\em repulsive} interaction. This 
somewhat counterintuitive effect has been observed in experiments with extensile microtubules and kinesin assemblies \cite{Sanchez:2012} and 
can be understood on the basis of the hydrodynamic approach embodied in Eqs. \eqref{eq:hydrodynamics}. In Fig. \ref{fig:pair_production} we 
have reproduced from Ref.~\cite{Sanchez:2012} a sequence of snapshots showing a pair of $\pm 1/2$ disclinations moving apart from each other 
together with the same behavior observed in our simulations.

\begin{figure}[t]
\includegraphics[width=\columnwidth]{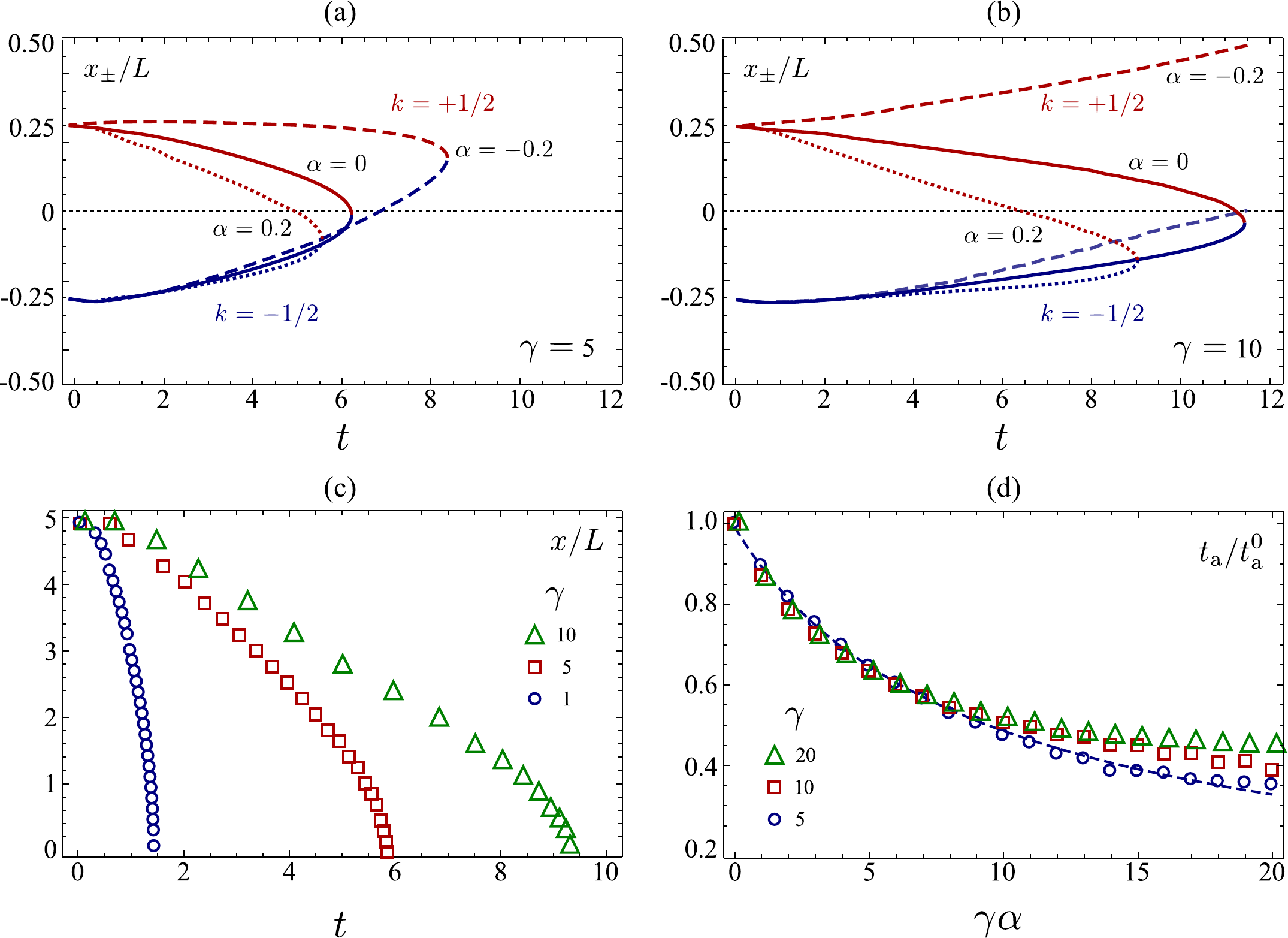}
\caption{(color online) Defect trajectories and annihilation times obtained from a numerical integration of Eqs. \eqref{eq:hydrodynamics} for 
various $\gamma$ and $\alpha$ values. (a) Defect trajectories for $\gamma=5$ and various $\alpha$ values (indicated in the plot). The upper 
(red) and lower (blue) curves correspond to the positive and negative disclination, respectively. The defects annihilate where the two curves 
merge. (b) The same plot for $\gamma=10$. Slowing down the relaxational dynamics of the nematic phase increases the annihilation time and for 
$\alpha=-0.2$ reverses the direction of motion of the $+1/2$ disclination. (c) Defect separation as a function of time for $\alpha=0.2$ and 
various $\gamma$ values. (d) Annihilation time normalized by the corresponding annihilation time obtained at $\alpha=0$ 
(i.e., $t_{\rm a}^{0}$). The line is a fit to the model described in the text. }
\label{fig:trajectories}
\end{figure}

To quantify the dynamics we have reconstructed the trajectories of the defects by tracking the drop in the magnitude of the order parameter. 
The trajectories are shown in Figs.~\ref{fig:trajectories}(a) and \ref{fig:trajectories}(b), where red lines in the upper portion of the plots
represent the trajectory of the $+1/2$ disclination, while the blue lines in the lower portion of the plot are the trajectories of the $-1/2$ 
defect. The tracks end when the cores of the two defects merge. For small activity and small values of the rotational friction $\gamma$, the 
trajectories resemble those obtained in Ref. \cite{Toth:2002} for passive systems. At large values of activity, however, the asymmetry in 
defect dynamics becomes more pronounced, and when the activity dominates over orientational relaxation, the $+1/2$ disclination moves 
independently along its symmetry axis with a velocity $\bm{v}\propto -\alpha\,\bm{\hat{x}}$, whose direction is dictated by the sign of 
$\alpha$. This behavior is clearly visible in Fig. \ref{fig:trajectories}(c), showing the defect separation $x(t)$ as a function of time. 
For $\gamma$ sufficiently large, the trajectories are characterized by two regimes. For large separation the dynamics is dominated by the 
active backflow, and thus $\dot{x}(t)\propto-\alpha$ and $x(t) \propto -\alpha t$. Once the defects are about to annihilate, the attractive 
force $F_{\rm pair} \propto 1/x$ takes over, and the defects behave as in the passive case with $x(t)\propto\sqrt{t_{\rm a}-t}$. 

Building on these results, we now propose a phenomenological one-dimensional model that captures qualitatively the dynamics of pair 
annihilation in active nematics. By neglecting for simplicity the position dependence of the friction, which we assume constant, the dynamics 
of a pair of disclinations initially at a distance $x_{0}$ along the $x$ axis is governed by the equations
\begin{equation}
\label{eq:xdot}
\mu\left[\dot{x}_{\pm}-v_{\rm b}(x_{\pm})\right]=\mp\frac{K}{x_+-x_-} \ ,
\end{equation}
where $v_{\rm b}(x_{\pm})$ is the backflow field at the position $x_\pm$ of the $\pm 1/2$ defect, given by 
$v_{\rm b}(x)=v_+(x-x_+)+v_-(x-x_-)$, with $v_\pm(x)$ the flow field due to an isolated $\pm 1/2$ defect. We retain only the active 
contribution to the backflow and replace the flow profiles by their constant values at the core of the defect, with 
$v_{\rm b}(x_{+})=v_{\alpha}\propto -\alpha$ and $v_{\rm b}(x_{-})=0$. Note that $v_{\alpha}<0$ for contractile systems and 
$v_{\alpha}>0$ for extensile ones. This yields the following simple equation for the pair separation
$\dot{x} = v_{\alpha} - \frac{2\kappa}{x}$, with $\kappa=K/\mu$. This equation explicitly captures the two regimes shown in 
Fig. \ref{fig:trajectories}(c) and described earlier. The solution takes the form
\begin{equation}\label{eq:x}
x(t)=x_0+v_{\alpha} t-2(\kappa/v_{\alpha})\ln\left[\frac{x(t)-2\kappa/v_{\alpha}}{x_0-2\kappa/v_{\alpha}}\right]\;.
\end{equation}
The pair annihilation time $t_a$ is determined by $x(t_a)=0$ and is given by 
$t_a=-x_0/v_\alpha-(2\kappa/v^{2}_{\alpha})\ln\left[1-(x_{0}v_{\alpha}/2\kappa)\right]$. For passive systems ($\alpha=0$) this reduces to 
$t_{a}^0=x_{0}^{2}/4\kappa$. In contractile systems, activity speeds up pair annihilation, while it slows it down in extensile systems. 
Our simple model predicts that the annihilation time, normalized to its value in passive systems, $t_a/t_a^0$, depends only on 
$v_\alpha x_{0}/2\kappa \sim \alpha\gamma$. Figure ~\ref{fig:trajectories}(d) shows a fit of the annihilation times extracted from the 
numerics to this simple formula. The model qualitatively captures the numerical behavior.  

\begin{figure}[t]
\includegraphics[width=\columnwidth]{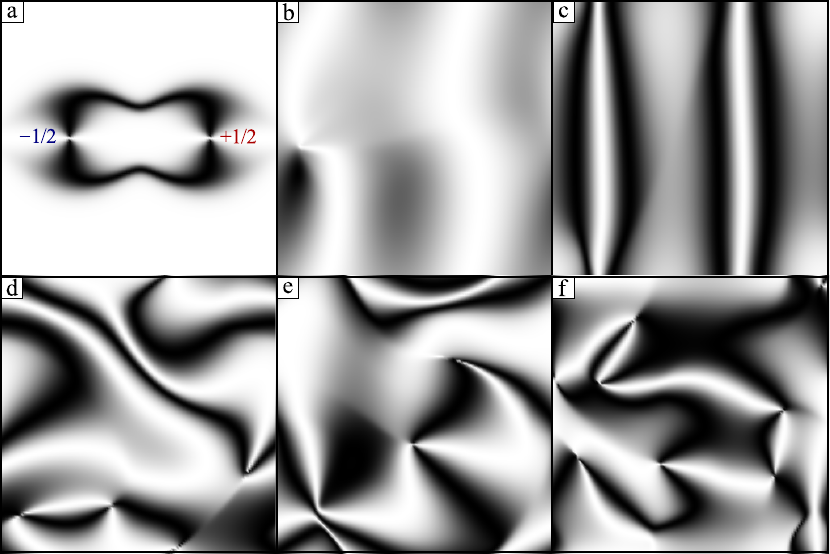}
\caption{\label{fig:post_collision}(color online) Schlieren texture highlighting the post-collisional dynamics of a $\pm 1/2$ pair for 
$\gamma=10^{3}$ and $\alpha=0.2$. (a) indicates the initial configuration of the defects, and (b) shows the system immediately after defect 
annihilation.}
\end{figure}

While the effect of activity on the precollisional dynamics of a disclination pair can be accounted for relatively simply in terms of  
active backflow, the postcollisional behavior is dramatically affected by activity \cite{Thampi:2013}. Figure ~\ref{fig:post_collision} 
shows the evolution of the system after annihilation of the initial defect pair (see also the supplementary movie S2~\cite{Supp}). 
The frame in Fig. \ref{fig:post_collision}(a) shows the initial configuration of the two defects, while Fig.\ref{fig:post_collision}(b) 
shows the configuration just after pair annihilation. The other frames display the evolution in time (with time increasing from left to 
right and top to bottom). Immediately after collision, the system develops two density or flow bands reminiscent of those observed in the
absence of defects~\cite{Giomi:2012}. The bands, however, are unstable and quickly start deforming while new defect pairs ``pinch off''. 
The dynamics quickly becomes chaotic, with frequent defect formation and annihilation events in the background of an 
overall proliferation of defects. The passage of defects through a region of space  lowers the local nematic order parameter in that region. 
At large friction $\gamma$, the slow relaxation prevents the restoration of the order parameter to its initial value, leading to a progressive
reduction of the average order parameter in time. Complex textures in active nematics were also reported in Ref. \cite{Fielding:2011}, 
although those structures are not easily decomposed in terms of disclinations. More work is needed to fully explore this rich and complex 
dynamics and formulate a quantitative classification of the behavior of defects in active liquid crystals.

We thank Zvonimir Dogic and Tim Sanchez for several illuminating discussions. L.G. was supported by SISSA mathLab. M.C.M. was supported by the
National Science Foundation through Grants No. DMR-1004789 and No. DGE-1068780. M.J.B. and X.M. were supported by the National Science 
Foundation through Grant No. DMR-0808812 and by funds from the Soft Matter Program of Syracuse University.


\begin{thebibliography}{1}

\bibitem{Marchetti:2013}
M. C. Marchetti, J. F. Joanny, S. R. Ramaswamy, T. B. Liverpool, J. Prost, M. Rao, and R. A. Simha, 
arXiv:11207.2929 [Rev. Mod. Phys. (to be published)]. 

\bibitem{Voituriez:2005}
R. Voituriez, J. F. Joanny and J. Prost, 
Europhys. Lett. {\bf 70}, 118102 (2005).

\bibitem{Marenduzzo:2007}
D. Marenduzzo, E. Orlandini, M. E. Cates, and J. M. Yeomans, 
Phys. Rev. E {\bf 76}, 031921 (2007).

\bibitem{Giomi:2008}
L. Giomi, M. C. Marchetti and T. B. Liverpool, 
Phys. Rev. Lett. {\bf 101}, 198101 (2008).

\bibitem{Ramaswamy:2003}
S. Ramaswamy, R. A. Simha and J. Toner, 
Europhys. Lett. {\bf 62}, 196 (2003).

\bibitem{Mishra:2006}
S. Mishra and S. Ramaswamy, 
Phys. Rev. Lett. {\bf 97}, 090602 (2006).

\bibitem{Narayan:2007}
V. Narayan, S. Ramaswamy, and N. Menon, 
Science 317, {\bf 105} (2007).

\bibitem{Sokolov:2009}
A. Sokolov and I. S. Aranson, 
Phys. Rev. Lett. {\bf 103}, 148101 (2009).

\bibitem{Giomi:2010}
L. Giomi, T. B. Liverpool and M. C. Marchetti, 
Phys. Rev. E {\bf 81}, 051908 (2010).

\bibitem{Fielding:2011}
S. M. Fielding, D. Marenduzzo, and M. E. Cates, 
Phys. Rev. E {\bf 83}, 041910 (2011).

\bibitem{Giomi:2011}
L. Giomi, L. Mahadevan, B. Chakraborty, and M. F. Hagan,
Phys. Rev. Lett. {\bf 106}, 218101 (2011).

\bibitem{Giomi:2012}
L. Giomi, L. Mahadevan, B. Chakraborty, and M. F. Hagan,
Nonlinearity {\bf 25}, 2245 (2012).

\bibitem{Wensink:2012}
H. H. Wensink, J. Dunkel, S. Heidenreich, K. Drescher, R. E. Goldstein, H. L\"owen, and J. M. Yeomans,
Proc. Natl. Acad. Sci. USA {\bf 109}, 14308 (2012).

\bibitem{Kleman:2003}
M. Kleman and O. Lavrentovich, 
{\em Soft Matter Physics: An Introduction} (Springer, New York, 2003).

\bibitem{Kemkemer:2000}
R. Kemkemer, D. Kling, D. Kaufmann, and  H. H. Gruler, 
Eur. Phys. J. E {\bf 1}, 215 (2000).

\bibitem{Sanchez:2012}
T. Sanchez, D. N. Chen, S. J. DeCamp, M. Heymann, and Z. Dogic, 
Nature (London) {\bf 491}, 431 (2012).

\bibitem{Kruse:2004}
K. Kruse, J. F. Joanny, F. J\"ulicher, J. Prost, and K. Sekimoto, 
Phys. Rev. Lett. {\bf 92}, 078101 (2004).

\bibitem{Kruse:2006}
K. Kruse and F. J. J\"ulicher, 
Eur. Phys. J. E {\bf 20}, 459 (2006).

\bibitem{Voituriez:2006}
R. Voituriez, J. F. Joanny, and J. Prost, 
Phys. Rev. Lett. {\bf 96}, 028102 (2006).

\bibitem{Elgetti:2011}
J. Elgeti, M. E. Cates, and D. Marenduzzo,
Soft Matter {\bf 7}, 3177 (2011).

\bibitem{DeGennes:1993}
P. G. de Gennes and J. Prost, 
{\em The Physics of Liquid Crystals} (Clarendon, Oxford, 1993).

\bibitem{Marchetti:2012}
M.C. Marchetti, Nature (London) {\bf 491}, 340 (2012).

\bibitem{foot}
The approximation of a two-dimensional system is justified in the experiments reported 
in \cite{Sanchez:2012} where the tracking of individual filaments does not show microtubules hopping over each other 
(Z. Dogic, private communication).

\bibitem{Denniston:1996}
C. Denniston, 
Phys. Rev. B {\bf 54}, 6272 (1996).

\bibitem{Toth:2002}
G. T\'oth, C. Denniston, and J. M. Yeomans, 
Phys. Rev. Lett. {\bf 88}, 105504 (2002).

\bibitem{Kats:2002}
E. I. Kats, V. V. Lebedev, and S. V. Malinin, 
J. Exp. Theor. Phys. {\bf 95}, 714 (2002).

\bibitem{Svensek:2002}
D. Sven\v{s}ek and S. \v{Z}umer,
Phys. Rev. E {\bf 66}, 021712 (2002).

\bibitem{Sonnet:2009}
A. M. Sonnet and E. G. Virga,
Liq. Cryst. {\bf 36}, 1185 (2009).

\bibitem{Pleiner:1988}
H. Pleiner, 
Phys. Rev. A {\bf 37}, 3986 (1988).

\bibitem{Ryskin:1991}
G. Ryskin and M. Kremenetsky, 
Phys. Rev. Lett. {\bf 67}, 1574 (1991).

\bibitem{Supp}
See supplemental Material at \url{http://link.aps.org/supplemental/10.1103/PhysRevLett.110.228101} for movies of disclination 
dynamics.

\bibitem{Thampi:2013}
S. P. Thampi, R. Golestanian and J. M. Yeomans,
arXiv:1302.6732 (2013).

\end{thebibliography}
\end{document}